\documentstyle[12pt,epsf]{article} 
\begin{document}
\def\singlespace {\smallskipamount=3.75pt plus1pt minus1pt
                  \medskipamount=7.5pt plus2pt minus2pt
                  \bigskipamount=15pt plus4pt minus4pt
                  \normalbaselineskip=12pt plus0pt minus0pt
                  \normallineskip=1pt \normallineskiplimit=0pt
                  \jot=3.75pt {\def\smallskip
                  {\vskip\smallskipamount}} {\def\medskip
                  {\vskip\medskipamount}} {\def\bigskip
                  {\vskip\bigskipamount}}
                  {\setbox\strutbox=\hbox{\vrule height10.5pt
                  depth4.5pt width 0pt}} \parskip 7.5pt
                  \normalbaselines} \def\middlespace
                  {\smallskipamount=5.625pt plus1.5pt minus1.5pt
                  \medskipamount=11.25pt plus3pt minus3pt
                  \bigskipamount=22.5pt plus6pt minus6pt
                  \normalbaselineskip=22.5pt plus0pt minus0pt
                  \normallineskip=1pt \normallineskiplimit=0pt
                  \jot=5.625pt {\def\smallskip
                  {\vskip\smallskipamount}} {\def\medskip
                  {\vskip\medskipamount}} {\def\bigskip
                  {\vskip\bigskipamount}}
                  {\setbox\strutbox=\hbox{\vrule height15.75pt
                  depth6.75pt width 0pt}} \parskip 11.25pt
                  \normalbaselines} \def\doublespace
                  {\smallskipamount=7.5pt plus2pt minus2pt
                  \medskipamount=15pt plus4pt minus4pt
                  \bigskipamount=30pt plus8pt minus8pt
                  \normalbaselineskip=30pt plus0pt minus0pt
                  \normallineskip=2pt \normallineskiplimit=0pt
                  \jot=7.5pt {\def\smallskip {\vskip\smallskipamount}}
                  {\def\medskip {\vskip\medskipamount}} {\def\bigskip
                  {\vskip\bigskipamount}}
                  {\setbox\strutbox=\hbox{\vrule height21.0pt
                  depth9.0pt width 0pt}} \parskip 15.0pt
                  \normalbaselines}

\begin{flushleft}
{\Large IC/96/43}
\end{flushleft}
\begin{center}
{\huge {Two-loop renormalization group analysis of supersymmetric
SO(10) models with an intermediate scale}} \\ \vskip 1in Mar
Bastero-Gil$\footnote{E-mail: bastero@stardust.sissa.it}^{a}$ and Biswajoy 
Brahmachari\footnote{ Address after 
1st Oct 96: Department of Physics and Astronomy, University of Maryland, 
College Park, MD-20742, USA. E-mail: biswajoy@ictp.trieste.it}$^{b}$ \\
\end{center}
\begin{center}
(a) Scuola Internazionale Superiore di Studi Avanzati \\ 34013
Trieste, ITALY. \\ (b) International Centre For Theoretical Physics,\\
34100 Trieste, ITALY.\\
\end{center}
\vskip 1in {
\begin{center}
\underbar{Abstract} \\
\end{center}
\middlespace
Two-loop evolutions of the gauge couplings in a class of intermediate
scale supersymmetric SO(10) models including the effect of third
generation Yukawa couplings are studied. The unification scale, the
intermediate scale and the value of the unification gauge coupling in
these models are calculated and the gauge boson mediated proton decay
rates are estimated. In some cases the predicted proton lifetime turns
out to be in the border-line of experimental limit. The predictions of
the top quark mass, the mass ratio $m_b(m_b)/ m_\tau(m_\tau)$
from the two-loop evolution of Yukawa couplings and the mass of the left 
handed neutrino via see-saw mechanism are summarized. The lower bounds on the
ratio of the VEVs of the two low energy doublets ($\tan\beta$) from
the requirement of the perturbative unitarity of the top quark Yukawa
coupling up to the grand unification scale are also presented. All the
predictions have been compared with those of the one-step unified
theory.}

\newcommand{\be}{\begin{equation}} \newcommand{\ee}{\end{equation}}
\newcommand{\bea}{\begin{eqnarray}} \newcommand{\eea}{\end{eqnarray}}
\def\tl{{\tilde{l}}} \def\tL{{\tilde{L}}} \def\bd{{\overline{d}}}
\def\tL{{\tilde{L}}} \def\a{\alpha} \def\b{\beta} \def\f{n_f}
\def\d{n_d} \def\h{n_H} \def\l{n_L} \def\r{n_R} \def\c{n_c}

\newpage
\doublespace
\section{Introduction and summary}
There is a general understanding in the literature that the LEP
measurements of the gauge couplings at the scale $M_Z$ and a
remarkable convergence of the couplings at the scale around $10^{16}$
GeV is a hint to a one step supersymmetric unified theory. There are
however several physical arguments suggesting that in a larger unified 
theory like SO(10) there may be an intermediate scale \cite{int1} 
corresponding to a left-right gauge symmetry breaking \cite{intscale} 
somewhere around $10^{11}$ to $10^{12}$ GeV based on
neutrino physics \cite{hdm,numass} as well as strong CP problem
\cite{kim}.  However, such a scale must not affect the gauge coupling
unification constraints. Recently a number of studies have been
performed to examine this question \cite{ sato, rizzo, leemoh,
biswamoha, vissani, bringole}. In this paper we present a systematic
two-loop analysis of a class of intermediate-scale supersymmetric SO(10)
models varying the value of the strong coupling 
constant
$\alpha_3(m_Z)$ in the experimentally allowed range. To begin with we
calculate the predictions of (i) unification scale $M_X$, (ii)
intermediate scale of B-L symmetry breaking $M_I$ (iii) the
unification coupling $\alpha_G(M_X)$ and (iv) the proton lifetime in a
two-loop renormalization group study.

The large value of the top quark mass measured by the CDF and DO
collaborations at FERMILAB \cite{topmass} is suggestive of a large top
Yukawa coupling.  Keeping this in mind we include the effect of all
the third generation Yukawa couplings in the running of the gauge
couplings. We note that in the large $\tan \beta$ region the effect of
the bottom quark and the tau lepton Yukawa couplings also affect the
gauge coupling evolution non-trivially, and one-step 
unification prefers a lower value of $\alpha_3(M_Z)=0.120$ instead of 
$0.124$ obtained in the low $\tan \beta$ region.
This effect also exists in the intermediate scale models. Later in 
this paper we focus our attention to the masses of fermions. We calculate
the (v) pole mass of the top quark when the top quark Yukawa coupling is at
the quasi-infrared fixed point at the scale $M_X$. The predictions of
(vi) the mass ratio $m_b(m_b)/ m_\tau(m_\tau)$ also emerges from
our analysis. The $\alpha_3$ dependence of these predictions are
displayed graphically and they are mostly consistent with the observed
values.

The ratio of VEVs $\tan \beta \equiv {\langle H^u \rangle /
\langle H^d \rangle }$ is an unknown parameter of the supersymmetric
models. However it has been noted that for very small values of $\tan
\beta$ the top quark Yukawa  coupling becomes very large before the scale
$M_X$ and consequently it is possible to give a lower bound on $\tan
\beta$ requiring the perturbative unitarity of the top
Yukawa coupling up to the scale $M_X$. In this paper we make an
analysis on the (vii) lower bound on $\tan \beta$ in such intermediate 
scale models. Taking the top
quark Yukawa coupling to be arbitrarily large at the scale $M_X$ we
derive an upper bound on it at the top scale 
[$h_t(m_t)$] and then we convert this upper bound to a lower bound on the
unknown parameter $\tan \beta$ for given values of the top quark mass.

We know that is difficult to understand the origin of a small 
neutrino-mass in the minimal supersymmetric standard model (MSSM) or in
the minimal SU(5) GUT. In the presence of a right handed singlet neutrino 
the VEV of the $SU(2)_L$ doublet Higgs scalar will provide a Dirac mass to 
the neutrino which comparable to the masses of the other fermions. On the 
other hand we also know that in an intermediate scale model such as 
SO(10) GUT, it is possible to relate the mass of the left-handed neutrino 
to the inverse of the large lepton number violating Majorana mass of the 
right-handed neutrino ($M_N \sim M_I$) by a see-saw mechanism 
\cite{numass}. We study the (viii) mass of the left-handed neutrino (of 
the third generation) in these models and calculate predictions for the 
tau neutrino mass by inputing of the values of the intermediate scales 
and various Yukawa couplings calculated from the first part of the paper. 
All our results can be compared with the one-step unification scenario 
in the limit $M_I=M_X$. 

We will consider the value of $\a_3(M_Z)$ \cite{pdg} in the range 0.110 
to 0.130, as the LEP measurements give $\alpha_3(M_Z)= 0.124\pm 0.006$
\cite{bethke}, which is typically 2 to 3 standard deviations larger than 
the value coming from $J/\Psi$--decay which is $\alpha_3(M_Z)= 0.108 \pm 
0.010$ \cite{kobel}, from lattice calculations which gives the value
$\alpha_3(M_Z)=0.110 \pm 0.006$ \cite{elkadra}, or from deep inelastic 
scattering experiments \cite{vir}which give $\a_3(M_Z)=0.112 \pm 0.002 \pm 
0.004$.

This paper is organized as follows. In section (II) we motivate our
models by a brief summary of an one-loop analysis. In
section (III) we give the two-loop analysis of the mass scales. In
section (IV) we give the predictions of top and the bottom quark
masses and derive lower bound on $\tan \beta$. In section (V) we
discuss the predictions of tau neutrino mass from see-saw mechanism and in 
section (VI) we conclude. The relevant formulae have been summarized in 
the appendix.

\section{Models and scalar contents: A case study} 

Before doing the rigorous analysis it is helpful to do a one-loop
case-study and fix the scalar contents of various models to be
considered afterwards. It is important to note for our purpose that an
unified model with $single$ intermediate scale is as predictive as a
model with grand desert in the following sense. Three low energy
observables in the gauge sector namely $\alpha_1(M_Z)$,
$\alpha_2(M_Z)$ and $\alpha_3(M_Z)$ can determine the three unknown
parameters namely $M_X$, $M_I$ and $\alpha_G(M_X)$. Given
$\alpha_1(M_Z)$ and $\alpha_2(M_Z)$ sufficiently precisely, there exists 
a unique value of $\a_3(M_Z)$ in the experimentally
allowed region for which one gets $M_X=M_I$. Indeed in a model with
more general scalar and fermion content one gets $M_X \ne M_I$, and 
thus, here we seek the models in which $M_I \le M_X$.

In one loop approximation this unique value, for which $M_X=M_I$, is 
$\alpha_3(M_Z) 
\sim 0.1144$. It is noted by Lee and Mohapatra \cite{leemoh} that it is
possible to get an intermediate scale $M_I < M_X $ in a number of models 
if $\alpha_3 \ne 0.1144$. In this paper one of our objectives is to find 
this unique value of $\alpha_3$ in two-loop approximation and derive the 
ranges for the intermediate scale for the allowed range of $\a_3(M_Z)$.
 
Now let us write down the 1-loop renormalization group equation 
(RGE)
of the three couplings introducing a general intermediate scale $M_I$
between $M_Z$ and $M_X$. We have used $b_i$ to denote the beta
function coefficients below the intermediate scale and $b_i^\prime$
to denote them above the intermediate scale. The relations are,
\begin{eqnarray}
\a_i^{-1}(M_Z)&=& \a_G^{-1}(M_X) + {b_i \over 2 \pi} \ln{M_I \over M_Z}
+ {b_i ^\prime \over  2 \pi} {M_X \over M_I}. \label{rgint}
\end{eqnarray}
A combination $\delta$ can be written \cite{bm1} in which $b_i$ get 
eliminated but $b_i^\prime$ survive, as, 
\be 
\delta =7 \a_3^{-1}(M_Z) - 12 \a_2^{-1}(M_Z) +
5 \a_1^{-1}(M_Z). \label{c2} 
\ee 
Eqn.(\ref{rgint}) and Eqn.(\ref{c2}) together leads to, 
\be \delta={ 1 \over 2 \pi} (7 b_3^\prime -12 b_2^\prime +5 b_1^\prime)
\ln {M_X \over M_I} \equiv {\Delta \over 2 \pi} \ln{M_X \over M_I} 
\label{c2b}. \ee
Now, if $\a_1(M_Z)=0.01696$, $\a_2(M_Z)=0.3371$ to get 
one-step unification in one loop we need $\a_3(M_Z)=0.1144$
which gives $ \delta=0$ from Eqn.(\ref{c2}). However, experimentally 
$\a_3(M_Z)$ can be in the range 0.11 to 0.13. Hence we obtain an allowed 
range in $\Delta$ which is crucial to get an intermediate scale, 
\be 
\delta^{min} \le { \Delta \over 2 \pi} \ln 
{M_X \over M_I}  \le \delta^{max}.  \label{cond}
\ee 
It may be noticed that when $\Delta$ is small we can achieve a 
large value for $\ln {M_X \over M_I}$ leading to an intermediate scale 
well-below the unification scale. We will return to the inequality 
\ref{cond} in a moment. Now let us consider the symmetry breaking pattern, 
\begin{eqnarray}
SO(10)&{M_X \atop \longrightarrow}& SU(3)_c \times SU(2)_L \times
SU(2)_R \times U(1)_{(B-L)}, \nonumber \\ &{M_I \atop \longrightarrow}&
SU(3)_c \times SU(2)_L \times U(1)_Y, \nonumber \\ &{M_Z \atop
\longrightarrow}& SU(3)_c \times U(1)_{em}. \nonumber
\end{eqnarray}
We restrict the type of Higgs representations above the intermediate
scale $M_I$ by requiring that the supersymmetric SO(10) GUT emerges
from an underlying superstring theory \cite{lykken}. Restricting
ourselves to only those Higgs scalars which can arise from simple
superstring models with Kac-Moody levels one or two, we can have a
restricted number of solutions of Eqn.(\ref{cond}). The various
models can be characterized by a set of five integers
($n_L,n_R,n_H,n_C,n_d$), where $n_C$ refers to the number of
(8,1,1,0), $n_H$ means the number of (1,2,2,0) fields and $n_L$ and
$n_R$ means the number of {(1,2,1,1)+(1,2,1,-1)} and
{(1,1,2,1)+(1,1,2,-1)} fields under the intermediate symmetry gauge
group and $n_d$ refers to the number of pairs of Higgs doublets below 
the scale $M_I$. To further restrict the number of models we have the 
following assumptions.

\noindent $\bullet$ The theory below the intermediate scale in strictly 
the Minimal Supersymmetric Standard Model (MSSM). We will also consider 
an example where the theory below $M_I$ has four Higgs doublets\footnote{
In this case the combination in the Eqn. (\ref{c2}) is 
no longer orthogonal to the beta function coefficients below $M_I$. 
However it is straight-forward to derive an equivalent combination 
for this case. Interesting low-energy phenomenology of such a 
model can be found in Ref. \cite{andrija}.} instead of two in the standard 
model.

\noindent $\bullet$ It should be possible to get an intermediate scale 
which is at least one to two orders of magnitude below the unification 
scale ( $\ln {M_X \over M_I}=2.3-4.6$).

Using the beta function coefficients listed in the appendix Eqn. 
(\ref{cond}) gives, \be
-10 \le (-9 + 21 n_c - 9 n_H + 6 n_R - 9 n_L) \le 6.
\label{ineq} 
\ee
The models satisfying the above inequality 
are tabulated in Table \ref{table2} which are numbered in the decreasing 
order of minimality. We have included one scenario namely model VIII 
which has a color octet 
above the scale $M_I$. This scenario has $\Delta=0$ and thus it can 
accommodate a large splitting between $M_I$ and $M_X$. In scenario  
IX the theory below the scale $M_I$ has four Higgs doublets 
($n_d=2$).
\begin{table}[htb]
\hfil \begin{tabular}{|c|| c c c c c||c|} 
\hline 
Model &$n_L$& $n_R$& $n_H$
 & $n_c$ & $n_d$ & $\Delta$ \\ 
\hline 
I & 0 & 2 & 1 & 0 & 1 & -6\\ 
II & 0 & 3 & 1 & 0& 1 &0\\
III & 0 & 4 & 1 & 0& 1 &6\\  
IV & 0 & 3 & 2 & 0 & 1 &-9\\ 
V & 0 & 4 & 2 & 0 & 1 &-3\\
VI & 0 & 5 & 2 & 0 & 1 & 3\\ 
VII & 1 & 5 & 2 & 0 & 1 &-6\\
\hline 
VIII& 1 & 1 & 1 & 1 &1 &0\\ 
\hline
IX&0&3&2&1&2&--\\
\hline
\end{tabular}
\hfil
\caption{The minimal models which satisfy the condition in Eqn.(5). 
When the quantity $\Delta$ is positive (negative) the model gives rise 
to an intermediate scale for the lower (higher) values of 
$\a_3(M_Z)$ than the one step unification case for which 
$\a_3(M_Z)=0.1144$ at the one-loop level. In the case $\Delta=0$ the 
intermediate scale is unconstrained at the one-loop level.} 
\label{table2} 
\end{table}

\section{Two loop analysis of mass scales, unified coupling and proton decay}
In the two-loop case the gauge couplings evolve according to the equation,  
\be 
{d \alpha_i \over d t} = {b_i \over 2 \pi}
\alpha^2_i + \sum_{j} {b_{ij} \over 8 \pi^2} \a^2_i \a_j -\sum_{k}
{a_{ik} \over 8 \pi^2} \a^2_i Y_k, \label{2lrg} 
\ee 
where the Yukawa
couplings can be defined by the superpotential invariant under the
intermediate symmetry, as,
\begin{eqnarray}
{\cal W_Y} &=& h_{Q_1} Q^T\tau_2\phi_1 Q^c + h_{Q_2} Q^T\tau_2\phi_2
Q^c + h_{L_1} L^T\tau_2 \phi_1 L^c + h_{L_2} L^T\tau_2\phi_2 L^c +
h.c.,
\end{eqnarray}
where we have denoted the quarks and leptons by the obvious notation
$Q, Q^c$ and $L,L^c$. At the GUT scale, we have
$h_{Q_1}(M_X)=h_{L_1}(M_X)=h_1(M_X)$ and
$h_{Q_2}(M_X)=h_{L_2}(M_X)=h_2(M_X)$.  $\phi_{1,2}$ are two Higgs
bidoublets embedded in $10_{1,2}$ scalars of SO(10). We have used the
notation, 
\be 
t= \ln \mu~~,~~\a={g^2 \over 4 \pi}~~,~~Y={h^2 \over 4
\pi}, 
\ee 
where g and h refers to the gauge and Yukawa couplings
respectively. The coefficients are given in detail in the
appendix. When $Y_1(M_X)$ and $Y_2(M_X)$ take large values of order
$O(1)$ the Yukawa couplings affect the running of the gauge couplings
most. In this case $Y_t$ and $Y_b$ have almost equal values at the low
energy scale. Such a situation corresponds to large $\tan \beta$. On
the other hand if $Y_2(M_X) << Y_1(M_X)$ at the scale $M_X$ the
top-bottom mass hierarchy is due to Yukawa coupling hierarchy
and consequently $\tan \beta$ is order O(1). We will
refer to this case as the small $\tan \beta$ scenario. In 
models I, II, III and VIII $\tan \beta$ is always large.

First, we present the results of the unification scale. The variation
of the unification scales with that of $\a_3(M_Z)$ have been plotted
in Figure (1a). The predictions of the intermediate scale have been
plotted in Figure (1b) and that of the unification coupling
$\alpha_G(M_X)$ have been plotted in Figure (1c) for various
models. The grand desert case can be recovered from the meeting point
of all the curves in Figure (1b) that is when the intermediate scale is
equal to the GUT scale. In the low $\tan \beta$ region the meeting
occurs at the value $\a_3(M_Z)=0.124$ and in the high $\tan \beta$
region it occurs at the value\footnote{ For moderate values of $\tan
\beta$ unification occurs for values of $\a_3(M_Z)$ within 0.120 and
0.124. Low energy threshold corrections \cite{threshold}assuming 
universality of gaugino masses will shift this whole range upwards.}
$\a_3(M_Z)=0.120$. For all other values of $\a_3(M_Z)$ it is possible
to have an intermediate scale. In model VI the unification scale
becomes low in the low $\a_3$ region; for model V the same thing happens 
for high $\a_3(M_Z)$ region.  As the dimension five 
proton decay can be suppressed in the SO(10) models by some additional
mechanism \cite{babr}, we plot the dominant dimension six decay mode
in Figure (1d).  The proton decay rate has been calculated using the
formula below in which we have taken the mass of the
heavy gauge bosons as $M_X$.  
\be \Gamma(p \longrightarrow e^+
\gamma) = {m_N \alpha^2\over 64 \pi f^2_\pi}~[{4 \pi \alpha_G~A_R
\over M^2_X}]^2~[1+ (1+ |V_{ud}|^2)^2] ~[1+D+F]. \label{pdecay} 
\ee
The values of parameters used in Eqn. (\ref{pdecay}), have been listed in 
the appendix.
\begin{figure}[t]
\begin{tabular}{cc}
\epsfysize=9cm \epsfxsize=9cm \hfil \epsfbox{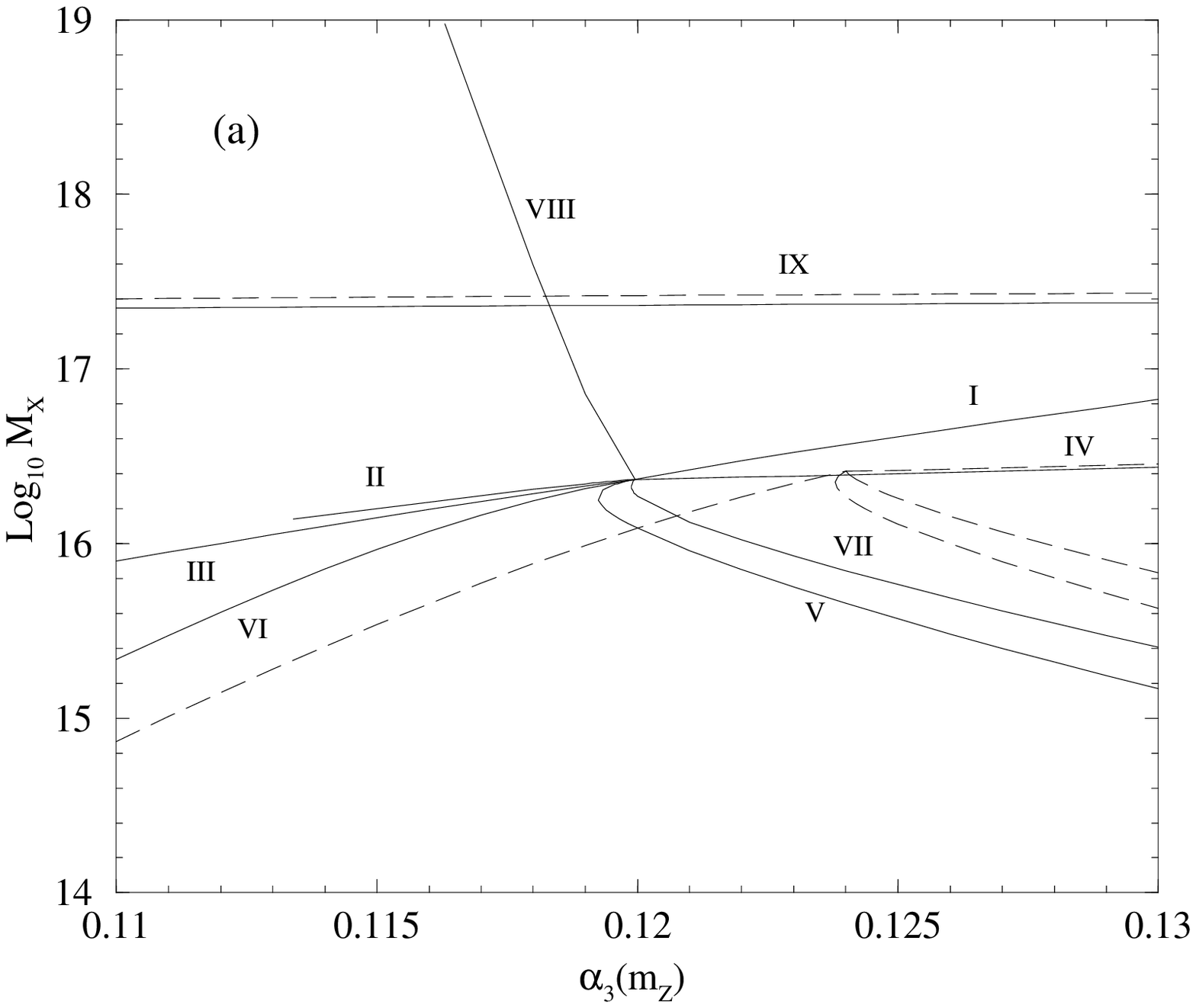} \hfil
& \epsfysize=9cm \epsfxsize=9cm \hfil \epsfbox{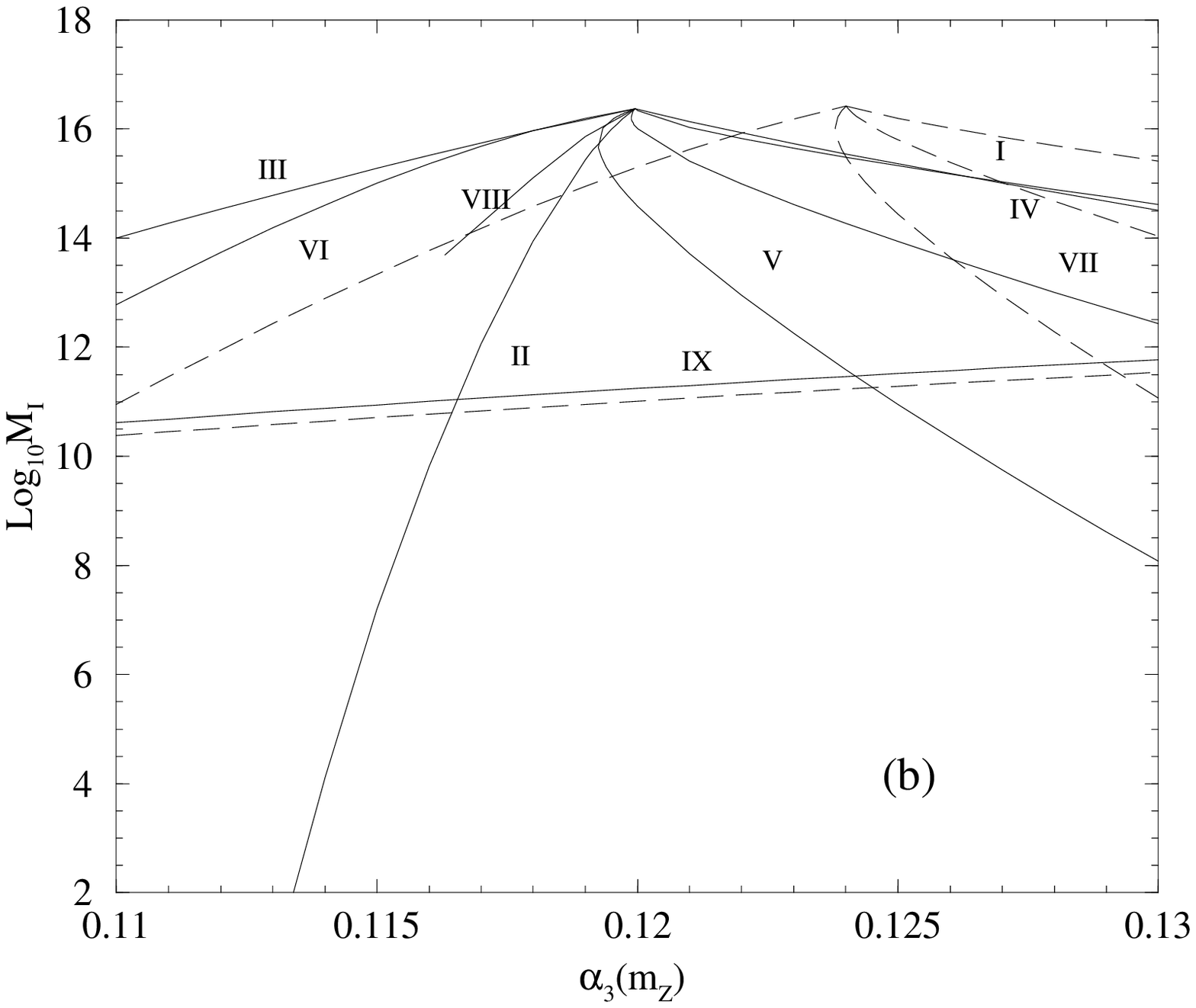} \hfil
\\ \epsfysize=9cm \epsfxsize=9cm \hfil \epsfbox{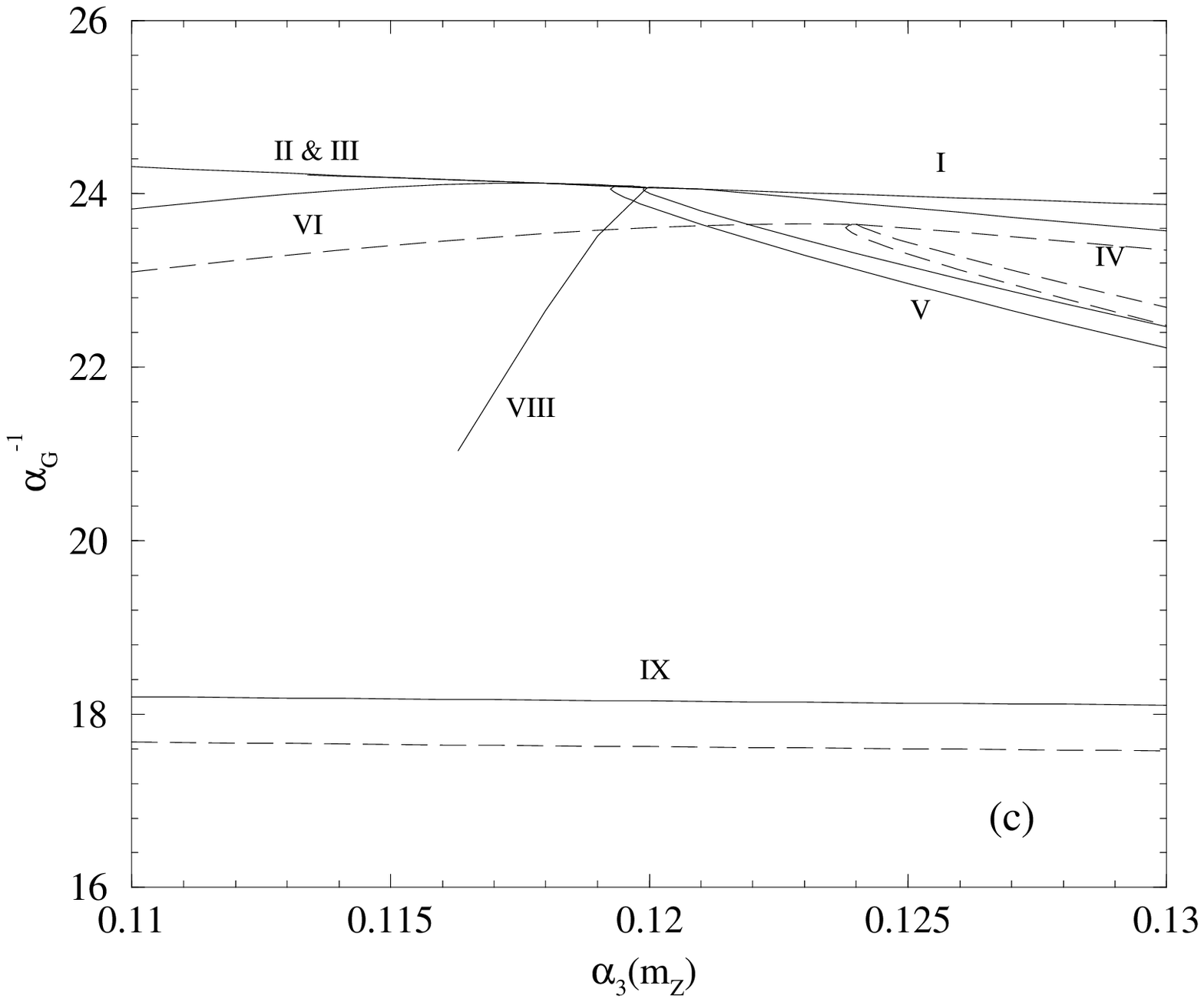} \hfil
& \epsfysize=9cm \epsfxsize=9cm \hfil \epsfbox{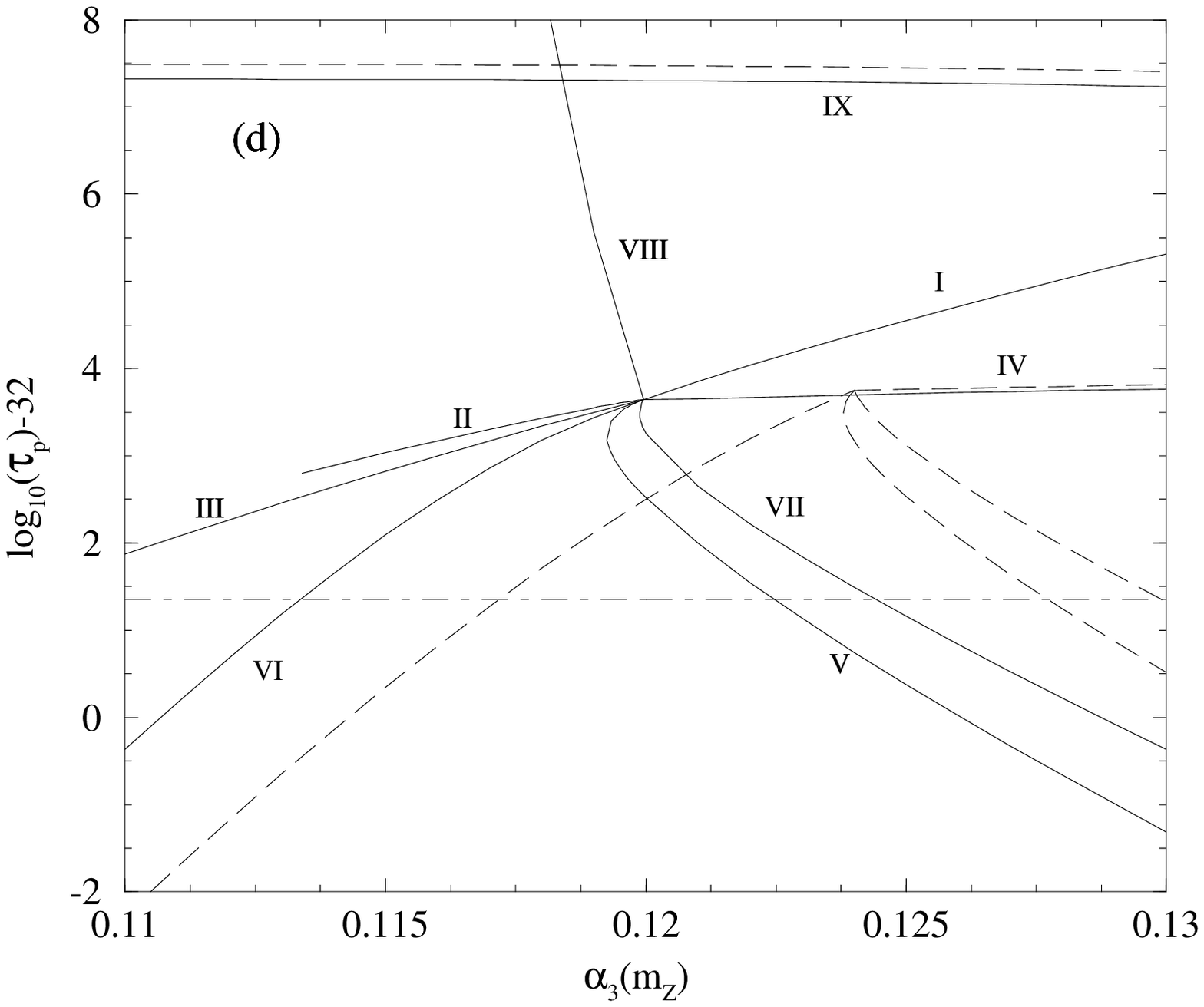} \hfil
\end{tabular}
\caption{ Predictions for (a) Unification scale, (b) Intermediate
scale, (c) Unification gauge coupling, and (d) Proton life--time, for
the models listed in Table I. Solid lines denote high $\tan \beta$
($Y_1(M_X)=Y_2(M_X)=1$; dashed lines denote the low $\tan \beta$
regime ($Y_1(M_X)=1$, $Y_2(M_X)=10^{-4}$). In Figure (d) the dotted
line is the experimental limit $\tau_p=5.5 \times 10^{32}\,yr$.}
\end{figure} 
The variation of decay mode vis-a-vis the variation of 
$\a_3(M_Z)$
comes not only from the variation of $M_X$ and $\alpha_G$ but also due
to a (mild) renormalization of the proton decay operator \cite{hisano,ar}
up to the scale of 1 GeV which has been parameterized in the factor
$A_R$. When $\alpha_3(M_Z)$ is above 0.130, $\alpha_3(1 GeV)$ becomes
very large and the perturbative calculation is not dependable. In
our estimation of  $A_R$ we have made taken into account the 
renormalization of the operator due to $SU(3)$ color and $SU(2)_L$ 
interactions only. This is a reasonable approximation as in most cases 
$M_I$ is large enough and the $SU(2)_R$ renormalization effects exist 
only beyond $M_I$. In the one-step unification case our value of the 
parameter $A_R$ agrees with that of Ref. \cite{ar}.

The lifetime of proton has been plotted in Figure (4). Model V
predicts large proton decay rates (in the borderline of experimental
limit) for the values of $\a_3(M_Z)$ beyond 0.123. Similarly model VI
predicts equally large rate of proton decay for lower values of
$\a_3(M_Z)$ below 0.114. The models except I, VIII and IX predicts a
lower proton lifetime than the minimal one step unification
scenario when there is an intermediate scale. This is encouraging from the 
forthcoming proton decay experiments where these models can be probed. In 
fact if the experiments at SuperKamiokande excludes the minimal unification
scenario it will also exclude all the present models except model I, VIII 
and IX.

Before concluding this section let us make a small remark about the 
curvatures visible in the plots near the meeting point of the models. 
These curvatures are more pronounced in models V and VII. Let us 
consider Figure (1b) for example. Near the GUT scale the Yukawa couplings 
are large and they fall quickly below the GUT scale. The Yukawa couplings 
tend to pull the individual lines towards the low $\a_3$ region wheres in 
the high $\a_3$ models (like V and VII) the gauge interactions have 
exactly the reverse effect. This causes the curvature in the graphs near 
the GUT scale which is purely a two-loop effect. 

\section{Fermion masses and Yukawa couplings}

The simultaneous evolution of gauge and Yukawa couplings enables us to
calculate the following Dirac type masses of quarks and leptons, in
respective scales denoted in the parenthesis, by the following
relations, 
\bea m_t(m_t)&=&\sqrt{4 \pi Y_t(m_t)} {v \over \sqrt{2}}
\sin \beta, \label{mt}
\\ 
m_b(m_b)&=& \sqrt{4 \pi Y_b(m_t)} {v \over
\sqrt{2}} \cos \beta ~\eta_b,
\label{mb}
\\
m_\tau(m_\tau)&=& \sqrt{ 4 \pi Y_\tau(m_t)} { v \over \sqrt{2}} \cos
\beta ~\eta_\tau. \label{mtau}
\\ 
m_{\nu_\tau}(M_I)&=& \sqrt{4 \pi
Y_{L1}(M_I)} {v \over \sqrt{2}} \sin \beta.  
\eea 
The factors $\eta_b$ and $\eta_\tau$ takes into account the running of 
the masses of $m_b$ and $m_\tau$ to their respective scales starting from 
$m_t$. We have used the formula of reference \cite{barger} to calculate 
$\eta_b$ and $\eta_\tau$ taking into account the three loop QCD effects 
and one loop QED effects. We have got $\eta_\tau=1.017$ assuming
$\alpha^{-1}_{em}(M_Z)=127.9$. The values of $\eta_b$ are plotted in
the appendix.
\begin{figure}[t]                                                        
\begin{tabular}{cc} 
\epsfysize=9cm \epsfxsize=9cm \hfil \epsfbox{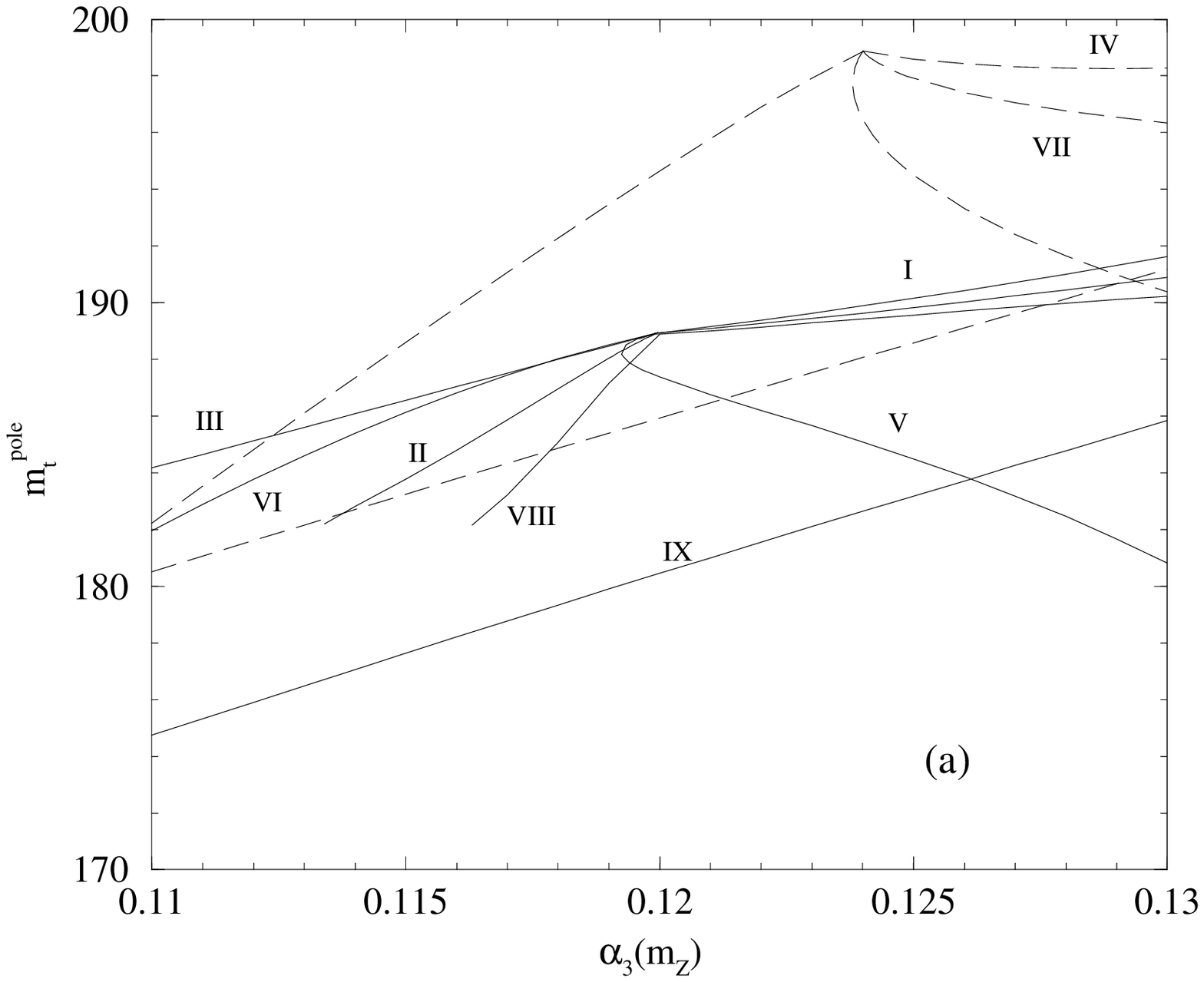} \hfil
&
\epsfysize=9cm \epsfxsize=9cm \hfil \epsfbox{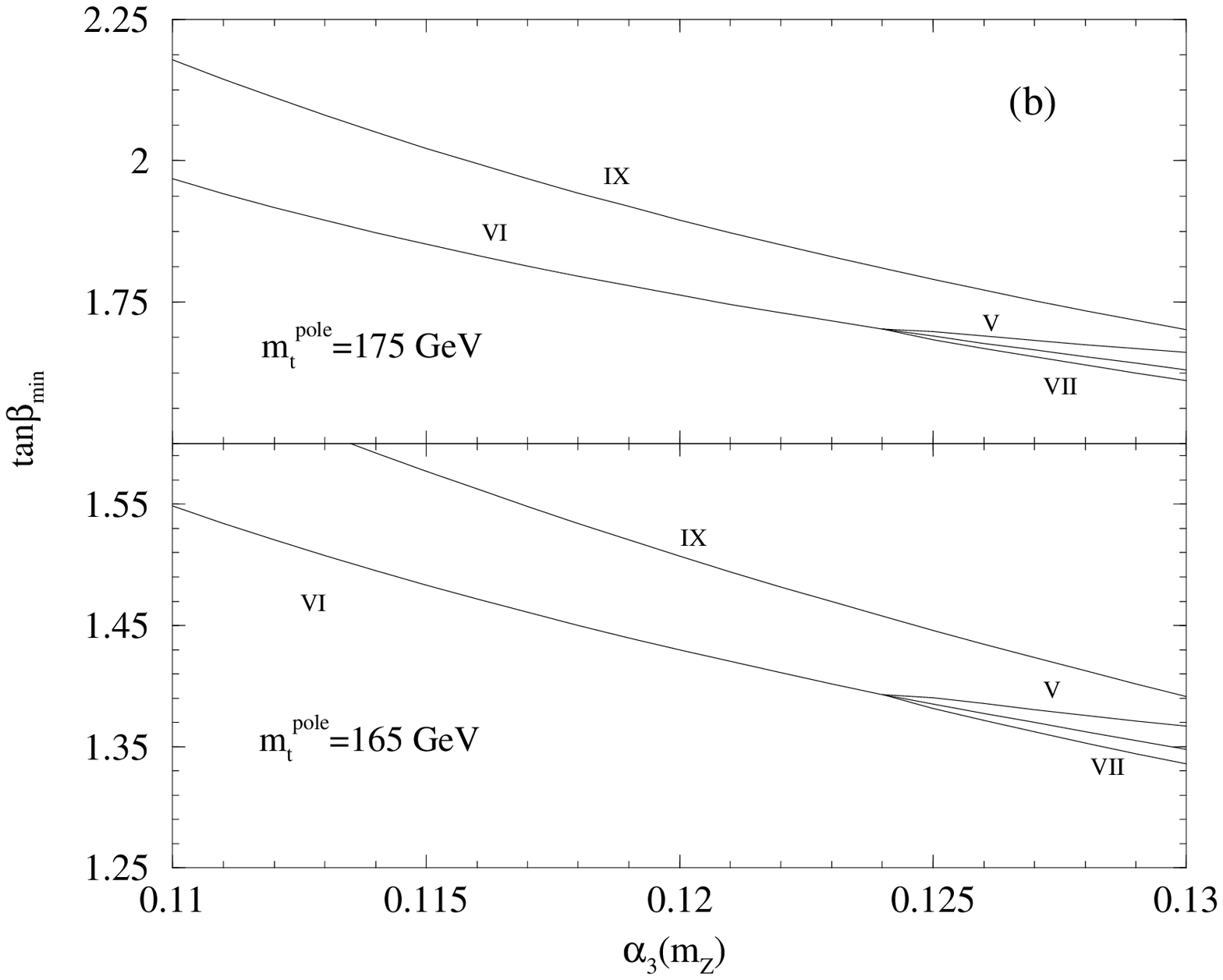} \hfil
\\
\epsfysize=9cm \epsfxsize=9cm \hfil \epsfbox{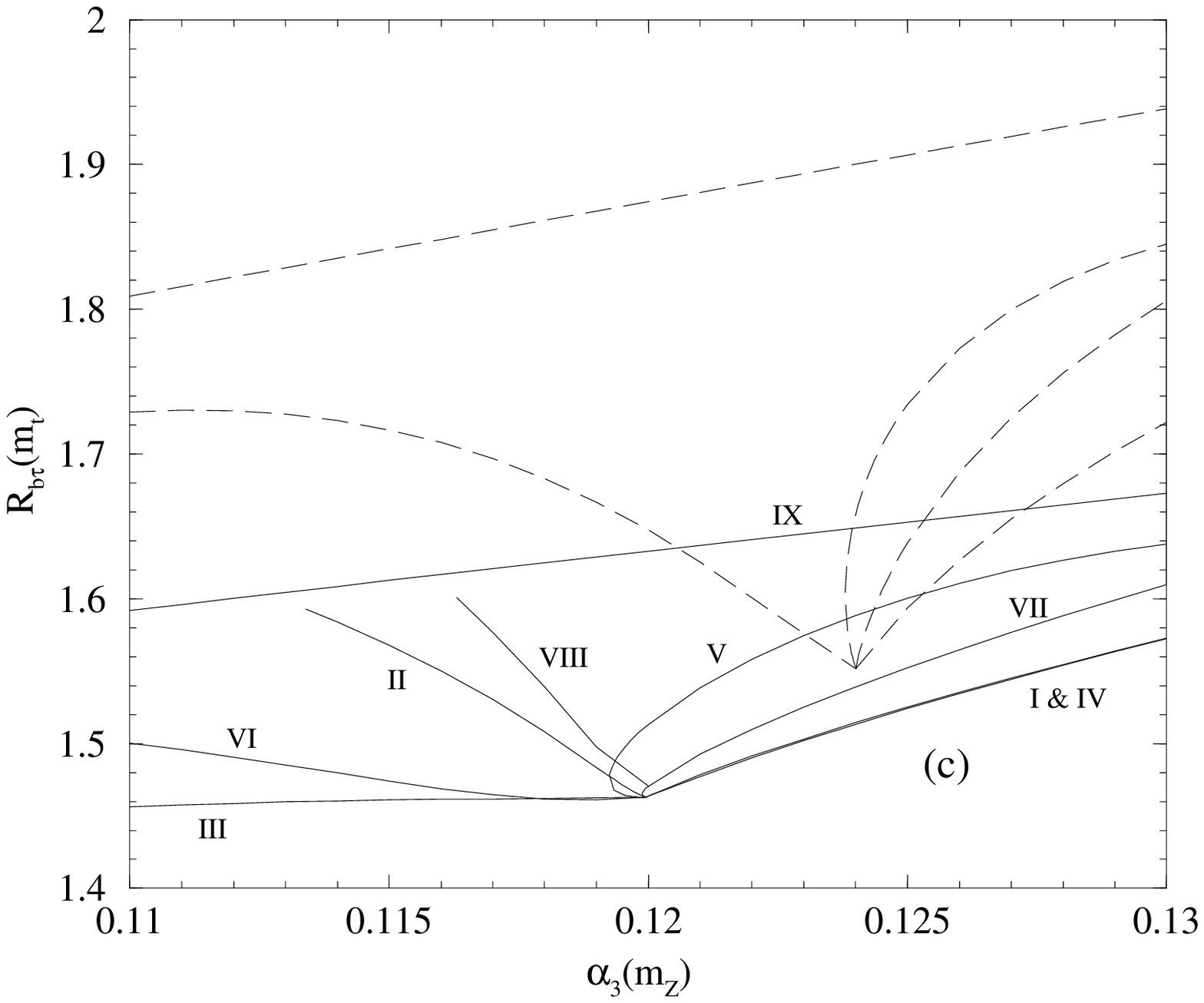} \hfil
&
\epsfysize=9cm \epsfxsize=9cm \hfil \epsfbox{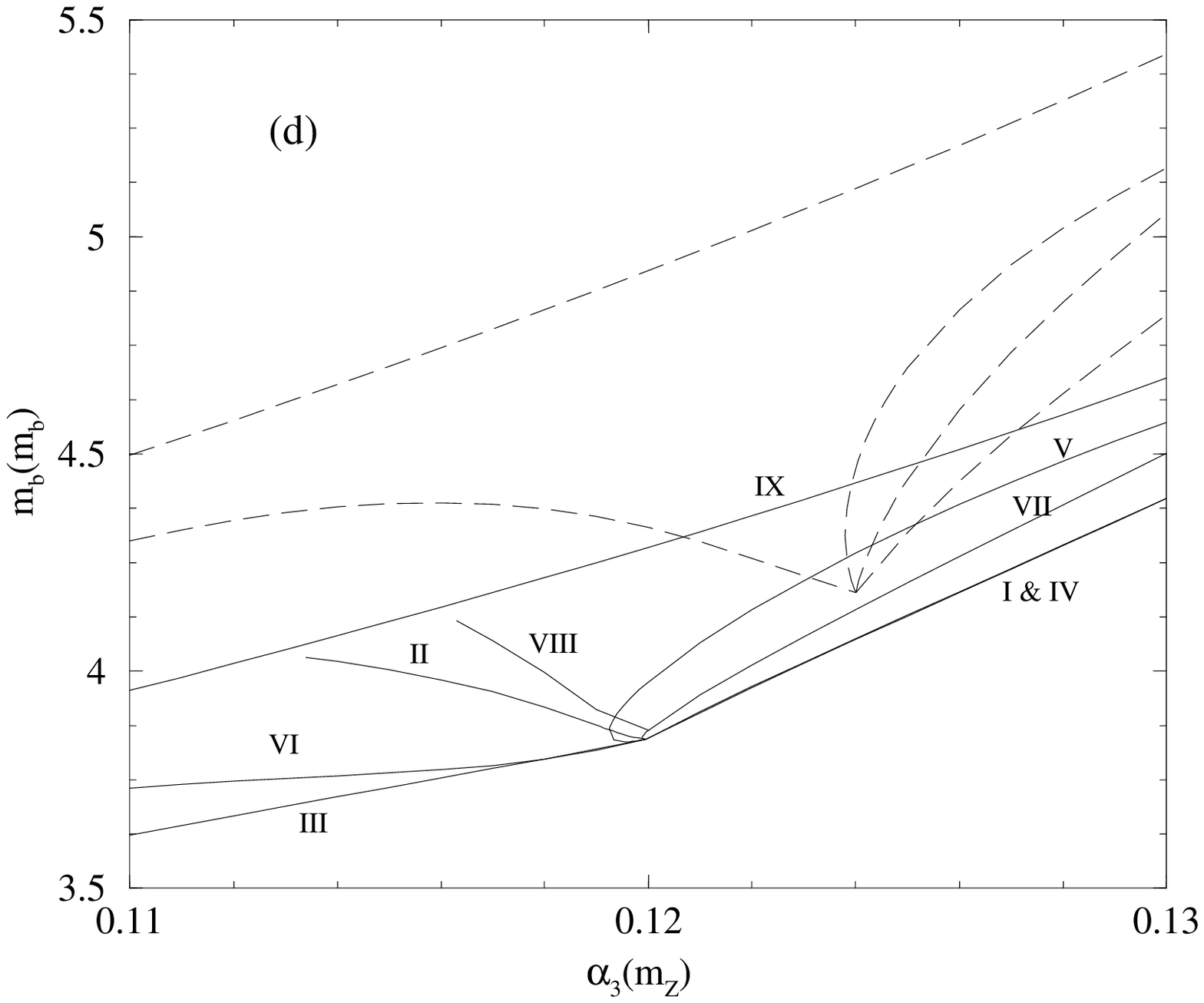} \hfil
\end{tabular}                                                                 
\caption{ Predictions of (a) pole mass of the top quark, (b) lower
bound on $\tan \beta$, (c) $R_{b\tau}$ at $m_t$, (d) running mass of
the b quark. Solid lines and dashed lines are as in Fig. (1).}  
\end{figure}                                                               
We start from a pair of values $Y_1(M_X)$ and $Y_2(M_X)$. Using the
RGE for the Yukawa couplings given in the appendix we calculate the
values of the couplings $Y_t(m_t)$, $Y_b(m_t)$ and $Y_\tau(m_t)$. At the 
scale $M_I$ the boundary conditions for the Yukawa couplings are,
\bea
Y_{Q1}~(M_I)&=&Y_t~(M_I)~~,
~~Y_{Q2}~(M_I)=Y_b~(M_I)~~\mbox{{\rm and}}~~Y_{L2}~(M_I) =Y_\tau~(M_I)
~~\mbox{{\rm when}}~~n_H=2,\\ 
Y_{Q}~(M_I)&=&Y_t~(M_I)=Y_b~(M_I)~~,                                            
~~\mbox{{\rm and}}~~Y_{L}~(M_I) =Y_\tau~(M_I)    
~~\mbox{{\rm when}}~~n_H=1.                                          
\eea
Using Equation Eqn. (\ref{mtau}), which has the least
experimental error we calculate the value of $\cos \beta$. Once the
value of $\cos \beta$ is known the predictions for $m_t(m_t)$ and
$m_b(m_b)$ follows from Eqn. (\ref{mt}) and Eqn. (\ref{mb}). The
running mass of the top quark has been calculated by iterative
procedure using the condition 
\be 
m_t(m_t)=m_t. \label{run} 
\ee 
The pole mass \cite{mtpole} has been calculated from the running mass in Eqn. 
(\ref{run}) using the  equation 
\be
m^{pole}_t=m_t(m_t)~[1 + {4 \a_3(m_t) \over 3 \pi}], \label{pole} 
\ee
for each value of $\alpha_3(M_Z)$. The predictions of $m^{pole}_t$ is
plotted in Figure (2a). The value of $R_b(m_t) \equiv {m_b(m_t) \over
m_\tau(m_t)}$ are plotted in Figure (2c), and when the prediction of
$m_b(m_t)$ is extrapolated to the mass scale of the bottom quark we
get Figure (2d) using $m_\tau(m_\tau)=1.777$ GeV.  A few comments are
in order.

\noindent $\bullet$ When we decrease $\sin \beta (\tan \beta)$, the
reduction in the bottom quark Yukawa effect increases the value of
$h_t(m_t)$. This is visible in the plot (2a). However, once the value of
$Y_2(M_X)$ is smaller than $10^{-4}$, the effect of the bottom quark
Yukawa coupling on $Y_t$ virtually ceases to exist. At this stage a
further reduction in $\sin \beta (\tan \beta)$ causes a large
reduction in the predicted value of the top mass. And in fact for
$Y_2(M_X)=10^{-5}$ the predicted top mass falls down to 161 GeV.

\noindent $\bullet$ The experimentally allowed range of $m_b(m_b)$ is
between 4.10 and 4.50 GeV \cite{pdg}. However masses as high as 5.2 GeV
are also considered in the literature. The predictions put strong
constraints on the model. Especially for the low $\a_3(M_Z)$ case, high
$\tan \beta$ regime is disfavored. In the low $\tan \beta$
scenario the predictions shift towards higher $\a_3(M_Z)$ which leads
to a bigger $\eta_b$ [see Fig.(7)] and hence a larger predicted value
of $m_b(m_b)$.

\noindent $\bullet$ The lower bounds on $\tan \beta (\sin \beta)$
follows directly from an upper bound on\footnote{It is well-known that
the Yukawa evolution equations have an infrared quasi-fixed point
structure \cite{cthill}.}  $Y_t(m_t)$. We calculate the bound by
the following procedure. Taking the value of $Y_t(M_X)$ to be
arbitrarily large we calculate the upper bounds for $Y_t(m_t)$ for
various $Y_2(M_X)$. The absolute upper bound is converted into a lower
bound for $\tan \beta$ by Eqn. (\ref{mt}) and Eqn. (\ref{pole}). These
bounds have been plotted in Fig.(2b).

\section{See-saw mechanism and neutrino mass}

If the neutrino is a Majorana particle it can have a lepton number
violating Majorana mass term. SO(10) GUT has a natural mechanism
to generate a large Majorana mass of the right handed neutrino. In
this case the neutrino mass matrix has the form, 
\be {\cal
M}=\pmatrix{0 & m^d\cr m^d & M}, 
\ee 
where $m^d$ is a $ 3 \times 3$
Dirac mass matrix and M is a $3 \times 3$ Majorana mass matrix in general. 
When the matrix ${\cal M}$ is diagonalized we get two eigenvalues of orders
$M$ and $[m^d]^2/M$. The latter eigenvalue can in principle explain
the smallness of the mass of the left-handed neutrino when M is large.

In our models the intermediate $B-L$ symmetry is broken by the Higgs
scalars $16+\overline{16}$ fields of SO(10).  We will consider two
different scenarios by which Majorana mass of the right handed
neutrino can be generated.

\noindent(a) Using a higher dimensional operator of the form ${h \over
M_X} 16_F 16_F 16_H 16_H$ written in terms of SO(10)
representations. The subscripts F and H mean fermions and scalars
respectively. When $16_H$ gets a VEV a large Majorana mass of the
order $h~v^2_R/M_X$ is generated.
  
\noindent(b) Introduction of additional singlets to have a generalized
see-saw mechanism \cite{leemoh,moha86}. In this case the mass is
effectively generated from a $3 \times 3$ see-saw matrix.  This case
is interesting as an effective $2 \times 2$ see-saw matrix
can be recovered purely from renormalizable interactions when the
singlet scalars get their VEVs.
 
The prediction of the mass of the left-handed neutrino in the case (a)
is well-known. We recast it in terms of the parameters being evaluated
by RGEs in our cases (we have used $M_I=g_R~v_R$), as, 
\be 
h~m_\nu(M_I)={ [h_{L1}(M_I) ~{v^u}]^2 \over v^2_R/M_X} = 8 \pi^2
\alpha_{2R} (M_I) {Y_{L1}(M_I)~[v^2 \tan^2 \beta /(1+\tan^2 \beta)]
\over M^2_I/M_X}. \label{massa} 
\ee 
In the case (b) \cite{leemoh,moha86}, the $3 \times 3$ mass matrix in the 
basis ($(\nu_a,N_a,S_a)$, where a=1,2,3 and $S_a$ are the singlets) is, 
\be
{\cal M}=\pmatrix{0&h_1 v^u&0 \cr h_1 v^u &0& h^\prime v_R \cr 0&
h^\prime v_R& M_s}, 
\ee 
where $M_s$ is the mass of the
singlet\footnote{For the detailed superpotential see
Ref\cite{leemoh}}. In this case the small mass of the left-handed
neutrino arises purely from the renormalizable interactions, as, \be
h^\prime ~m_\nu(M_I)={ [h_{L1}(M_I) ~{v^u}]^2 \over v^2_R/M_s} = 8
\pi^2 \alpha_{2R} (M_I) {Y_{L1}(M_I)~[v^2 \tan^2 \beta /(1+\tan^2
\beta)] \over M^2_I/M_s}. \label{massb} \ee In Eqn. (\ref{massb}) we
will consider $M_s=M_I$. When $M_s=M_X$ we recover Eqn. (\ref{massa})
from Eqn. (\ref{massb}).  In both the cases we will assume, 
\be
v^2=(v^u)^2+(v^d)^2=247^2/2~GeV^2.  
\ee
\begin{figure}[t]                                                        
\begin{tabular}{cc} 
\epsfysize=9cm \epsfxsize=9cm \hfil \epsfbox{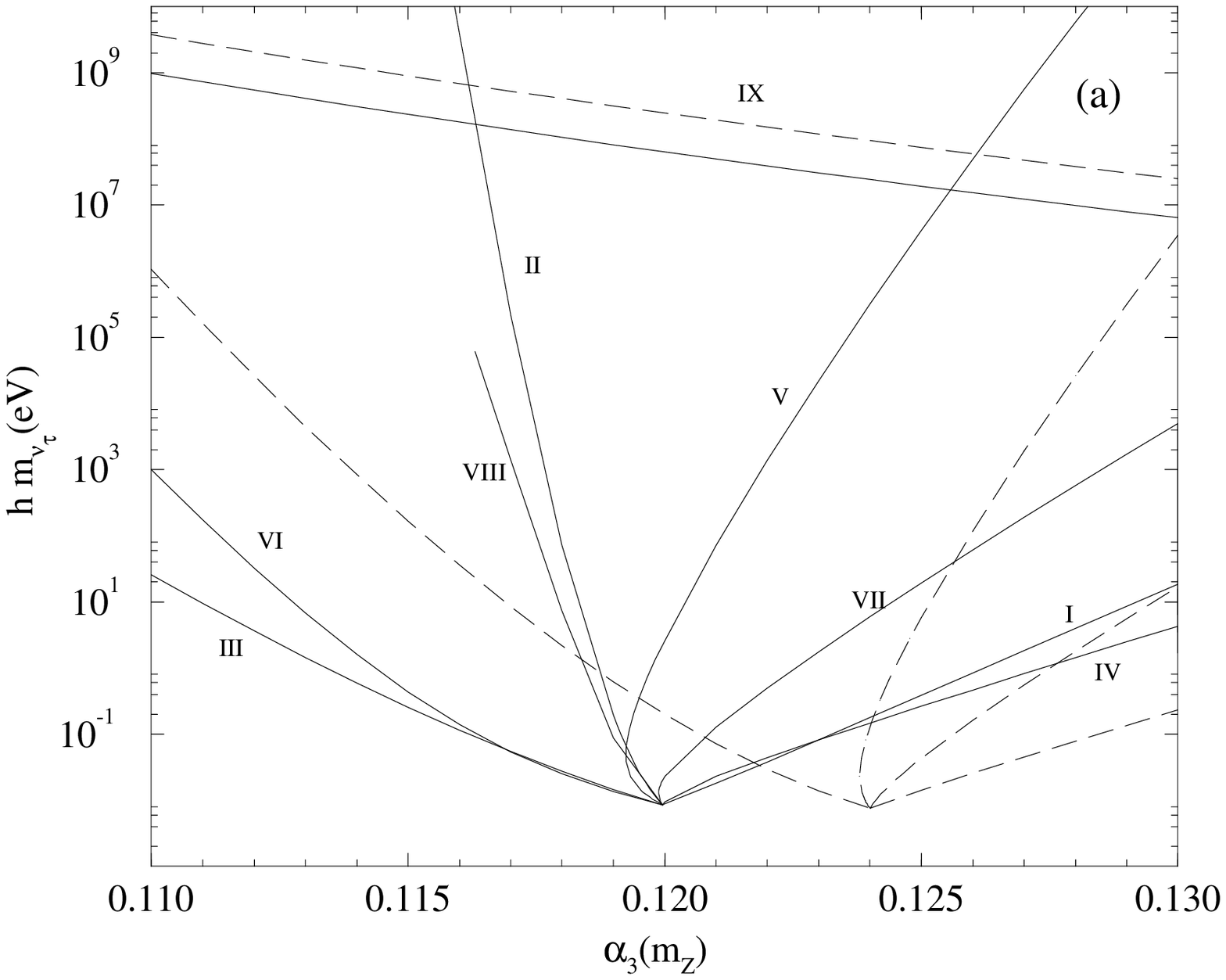} \hfil
&
\epsfysize=9cm \epsfxsize=9cm \hfil \epsfbox{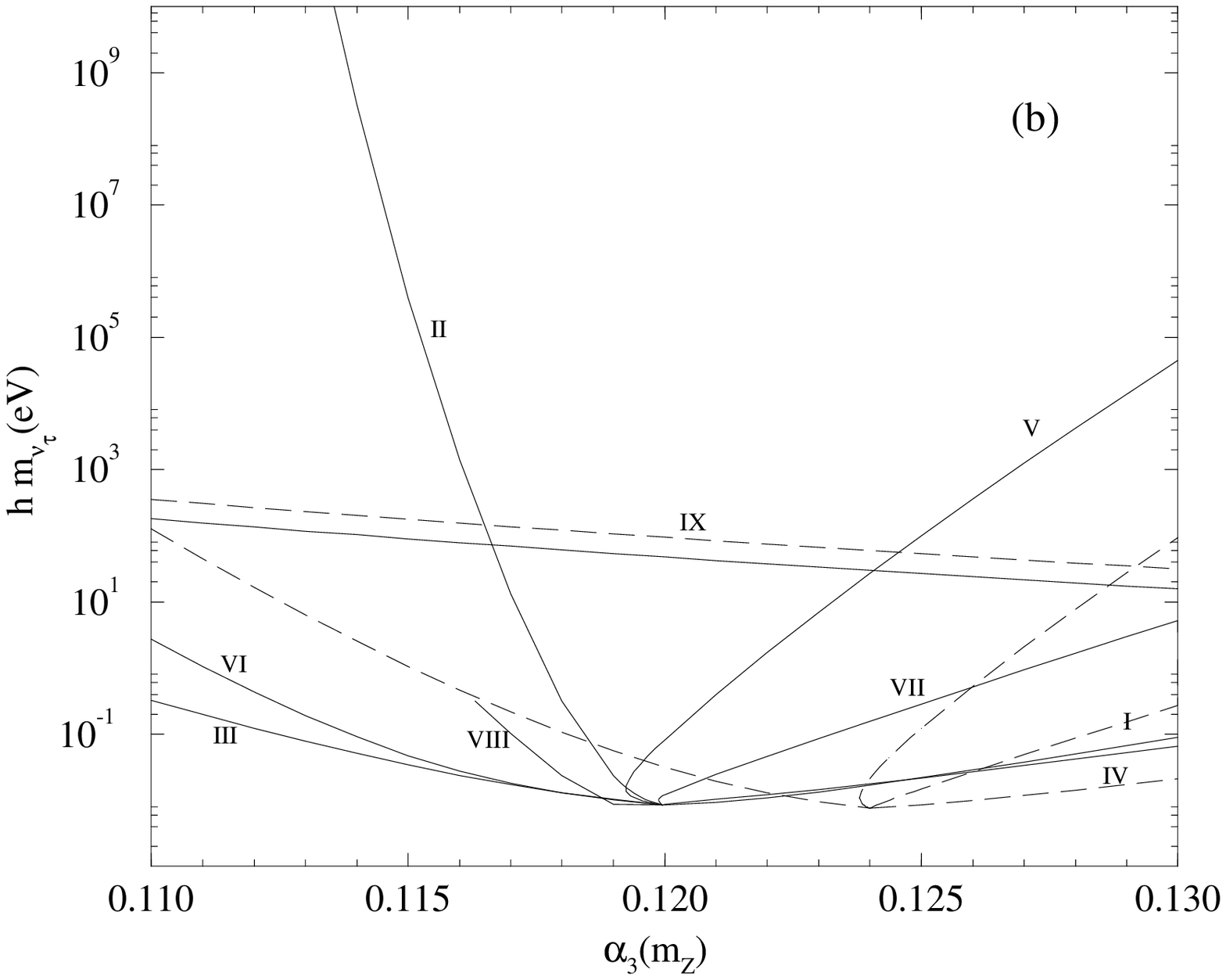} \hfil
\end{tabular} 
\caption{ Predictions of the left--handed neutrino mass by see--saw
mechanism by the two scenarios (a) and (b). Solid and dashed lines are
as before.}
\end{figure}
In Figure (3a) and Figure (3b) we have plotted the left handed neutrino
masses in scenarios (a) and (b) modulo the unknown Yukawa couplings $h$ and 
$h^\prime$ in various models as a function of $\a_3(M_Z)$. These numbers 
will get renormalized when they are extrapolated up to the scale of 1 GeV
\cite{bludman, babu, parida}. In Ref \cite{parida} it has been noted 
that the Yukawa couplings affect the extrapolation of the see-saw 
formula and the tau neutrino mass increases. This extrapolation will 
change the mass of the left handed neutrino by a factor of the order of 
unity.

We know that a tau neutrino mass of the order of a few electron volts
is preferable if neutrino is to be a candidate for the Hot Dark Matter
(HDM). We see that in scenario (a) a tau neutrino in the range of
1-10 eV can be achieved in all the models depending on the value of
$\a_3(M_Z)$. On the other hand, scenario (b) can predict a tau
neutrino mass in the 1-10 eV range for models V, II and VI. However,
model VI and V have potential problems with proton decay around the
values of $\a_3(M_Z)$ needed to produce a correct order of the mass of
the $\nu_\tau$, whereas model II predicts a lower value for the 
bottom quark mass.

\section{Conclusions}

We have considered a class of minimal models for which an intermediate scale
in a SO(10) GUT can be achieved. We have found that a small number 
$\Delta$ can be defined, the sign of which can tell whether the unification
occurs for a low value of $\alpha_3$ or a high value of $\alpha_3$ 
compared to the one-step unification. After listing the minimal models in 
which an intermediate scale can be generated, we have done a two-loop 
analysis of the gauge couplings to find 
the unification scale, the intermediate scale and the unification gauge 
coupling. These predictions are in general functions of the input 
$\alpha_3(M_Z)$. We have plotted the predictions in Figure 1. The 
predictions for one-step unification can be recovered from the meeting 
point of all the branches in these figures for which $M_I=M_X$. The gauge 
boson mediated proton decay rates have been calculated after 
renormalizing the proton decay operator to the scale of $1GeV$. These 
predictions are also plotted in the Figure 1. In some models the 
predicted proton life-time is in the borderline of experimental exclusion 
limits. These models can be tested in the SuperKamiokande experiments 
soon.

We have also done a parallel two-loop analysis of the Yukawa sector of 
these models. The combined running of the gauge and Yukawa couplings 
enable us to calculate the predictions of the top and bottom quark masses 
using the mass of the tau lepton $m_\tau(m_\tau)$ as an input. The 
predictions of the top mass is always in the range given by the CDF and 
D0 collaborations. The bottom quark mass is not always in the range 
4.10-4.50 GeV as quoted in the review of particle properties 
\cite{pdg}. Thus predictions of the bottom quark mass is a good indicator 
by which one could compare these models relative to each other. 
Running of the Yukawa coupling also gives a lower bound on the parameter 
$\tan \beta$ from the requirement that the top quark Yukawa coupling 
remains perturbative up to the scale of unification.

The determination of the intermediate scale and the running of the Yukawa 
couplings enable us to calculate the predictions of the neutrino masses 
via see-saw mechanism. The Dirac type entry in the see-saw matrix can be 
fixed from the running of the neutrino Yukawa coupling. The predictions 
of the tau neutrino for various values of $\alpha_3$ have been plotted. A 
mass of the tau neutrino in the range 1-10 electron volts can be achieved 
in various models. It can be noted from the plot of the neutrino masses 
that in the one-step unification model the mass of the tau neutrino is in 
the range of $10^{-2}$ electron volts only, making it less attractive as 
a dark matter candidate.

The scenario VIII can be a good candidate for string unification with the 
unification scale near the plank scale and the unification coupling 
larger than the one-step unification case. 

\section*{Acknowledgments}
We thank R. N. Mohapatra and G. Senjanovi\'c for useful comments. We also 
thank J. Kubo for sending us a numerical subroutine which we have 
used in our analysis.

\section{Appendix}
\subsection{Beta function coefficients for the Yukawa couplings}
It is easy to write down the
evolution equations for the Yukawa couplings defined by  $\sqrt{4 \pi
Y_{ijk}}~ \phi_i \phi_j \phi_k$, as,
\be
{d Y_{ijk} \over d t}=2 Y_{ijk}~[\gamma_i + \gamma_j + \gamma_k].
\ee
The scale t is defined as $t=\ln \mu$. In the two loop approximation we 
can write, \be   
\gamma_i=\gamma^{(1)}_i+\gamma^{(2)}_i.
\ee
The superscripts in parenthesis 
denote the order of perturbation theory. 
The anomalous dimensions for the superfields below the scale $M_I$ are 
given below. In the one-loop anomalous dimensions are,
\bea
\gamma^{(1)}_L& =& { 1 \over 4 \pi}~[ Y_\tau 
- { 3 \over 2} \a_2 - { 3 \over 10} \a_1 ]\\ 
\gamma^{(1)}_{\overline{E}}&=& { 1 \over 4 \pi }~[2 
Y_\tau - { 6 \over 5} \a_1],  \\ 
\gamma^{(1)}_{\overline{D}}&=& 
{ 1 \over 4 \pi^2}~[2 Y_b-{ 8 \over 3} \a_3 - { 4 \over 30} \a_1], \\
\gamma^{(1)}_{\overline{U}}&=&{ 1\over 4 \pi}~[ 2 Y_t-{ 8 \over 3} \a_3 - 
{ 8 \over 15} \a_1], \\ 
\gamma^{(1)}_Q&=& { 1 \over 4 \pi}~[ Y_t + Y_b - { 8 \over 3} \a_3 - { 3 
\over 2} \a_2 - { 1 \over 30} \a_1],\\
\gamma^{(1)}_{H_1}&=& { 1 \over 4 \pi}~[Y_\tau+ 3 Y_b - { 3 \over 2} 
\a_2 - { 3 \over 10} \a_1],\\
\gamma^{(1)}_{H_2}&=& { 1 \over 4 \pi}~[3 Y_t - { 3 \over 2} \a_2 - { 
3 \over 10} \a_1].
\eea
The two-loop anomalous dimensions are;
\bea                                
\gamma^{(2)}_L& =& { 1 \over 16 \pi^2}~[ -Y_\tau~ 
(\gamma^{(1)}_{\overline{E}} + \gamma^{(1)}_{H_1})\nonumber\\ 
&&- ({ 3 \over 2} \a_2 + { 3 \over 10} \a_1) \gamma^{(1)}_L
+ { 3 \over 2} b_2 \a^2_2 + { 3 \over 10} b_1 \a^2_1] , \\ 
\gamma^{(2)}_{\overline{E}}&=& { 1 \over 16 \pi^2 }~[ 
-2 Y_\tau~(\gamma^{(1)}_L + \gamma^{(1)}_{H_1}) 
\nonumber\\                    
&&- { 6 \over 5} \a_1 \gamma^{(1)}_{\overline{E}} 
+ { 6 \over 5} b_1\a^2_1]  
, \\                                           
\gamma^{(2)}_{\overline{D}}&=& 
{ 1 \over 16 \pi^2}~[-2 Y_b ~(\gamma^{(1)}_Q + \gamma^{(1)}_{H_1}) 
\nonumber\\ 
&&-({ 8 \over 3} \a_3 + { 2 \over 15} \a_1) \gamma^{(1)}_{\overline{D}}
+{ 8 \over 3} b_3 \a^2_3 + { 2 \over 15} b_1 \a^2_1] , \\ 
\gamma^{(2)}_{\overline{U}}&=&{ 1\over 16 \pi^2}~[ -2 Y_t~
(\gamma^{(1)}_{Q} + \gamma^{(1)}_{H_2}) \nonumber\\
&&-( { 8 \over 3} \a_3 - { 8 \over 15} \a_1) \gamma^{(1)}_{\overline{U}}
+ { 8 \over 3} b_3 \a^2_3 + { 8 \over 15} b_1 \a^2_1
] 
, \\                                 
\gamma^{(2)}_Q&=& { 1 \over 4 \pi}~[ -Y_t ~
(\gamma^{(1)}_{\overline{U}} + \gamma^{(1)}_{H_2})
- Y_b~(\gamma^{(1)}_{\overline{D}} + \gamma^{(1)}_{H_1}) \nonumber\\ 
&&- ({ 8 \over 3} \a_3 + { 3 \over 2} \a_2 + { 1 \over 30} 
b_1 \a_1) 
\gamma^{(1)}_Q
+ { 8 \over 3} b_3 \a^2_3 + { 3 \over 2} b_2 \a^2_2 + { 1 \over 
30} b_1 \a^2_1
], \\                          
\gamma^{(2)}_{H_1}&=& { 1 \over 16 \pi^2}~[-Y_\tau
~(\gamma^{(1)}_L + \gamma^{(1)}_{\overline{E}})
- 3 Y_b~ 
(\gamma^{(1)}_Q + \gamma^{(1)}_{\overline{D}}) \nonumber\\
&&- ({ 3 \over 2} \a_2 + { 3 \over 10} \a_1) \gamma^{(1)}_Q
+ { 3 \over 2} b_2 \a^2_2 + { 3 \over 10} b_1 \a^2_1
], \\                              
\gamma^{(2)}_{H_2}&=& { 1 \over 16 \pi^2}~[-3 Y_t ~
(\gamma^{(1)}_Q + \gamma^{(1)}_{\overline{U}}) \nonumber\\
&&- ({ 3 \over 2} \a_2 + { 3 \over 10} \a_1) \gamma^{(1)}_{H_2}
+{ 3 \over 2} b_2 \a^2_2 + { 3 \over 10} b_1 \a^2_1
].                         
\eea                                                          
Above the scale $M_I$ the one-loop anomalous dimensions are, 
\bea
\gamma^{(1)}_Q&=& {1 \over 4~\pi} [ 2Y_{Q1}
+2 Y_{Q2} - 
{8 \over 3} \alpha_c-{3 \over 2} \alpha_L - { 1 \over 12} \alpha_{B-L}] , \\ 
\gamma^{(1)}_{Q^c} &=& { 1 \over 4 
\pi} [2 Y_{Q1} + 2 Y_{Q2}  - { 8 \over 3} \alpha_c - { 3 \over 2} \alpha_R 
- { 1\over 12 } \alpha_{B-L}] , \\
\gamma^{(1)}_L &= & { 1\over 4 \pi} [2 Y_{L1} 
+ 2 Y_{L2} - { 3 \over 2} \alpha_L - { 3 \over 4} \alpha_{B-L} ] , \\
\gamma^{(1)}_{L^c} &=& { 1 \over 4 \pi 
} [ 2 Y_{L1} + 2 Y_{L2} - { 3 \over 2} \alpha_R - { 3 \over 4} \alpha 
_{B-L} ], \\
\gamma^{(1)}_{\phi_i} & =& { 1 \over 4 \pi}[ 3 Y_{Qi} + Y_{Li}  -{3 \over 2} 
\alpha_L - {3 \over 2} \alpha_R] .
\eea
The two-loop anomalous dimensions are,
\bea
\gamma^{(2)}_Q&=& {1 \over 16~\pi^2} [ -2 Y_{Q1}~(\gamma^{(1)}_{\phi_1}+ 
\gamma^{(1)}_{Q^c})  
- 2 Y_{Q2}~(\gamma^{(1)}_{\phi_2}+ \gamma^{(1)}_{Q^c}) \nonumber \\
&&-({8 \over 3} \alpha_c+{3 
\over 2} \alpha_L + { 1 \over 12} \alpha_{B-L}) \gamma^{(1)}_Q+
{8 \over 3} b_3 \alpha^2_3 + {3 \over 2} b_L \alpha^2_L + { 1 \over 12} 
b_{B-L} \alpha^2_{B-L}] 
, \\
\gamma^{(2)}_{Q^c}&=& {1 \over 16~\pi^2} [ -2 Y_{Q1}~(\gamma^{(1)}_{\phi_1}+
\gamma^{(1)}_{Q})
- 2 Y_{Q2}~(\gamma^{(1)}_{\phi_2}+ \gamma^{(1)}_{Q}) \nonumber \\
&&-({8 \over 3} \alpha_c+{3
\over 2} \alpha_R + { 1 \over 12} \alpha_{B-L}) \gamma^{(1)}_{Q^c}+
{8 \over 3} b_3 \alpha^2_3 + {3 \over 2} b_R \alpha^2_R + { 1 \over 12}
b_{B-L} \alpha^2_{B-L}] 
, \\
\gamma^{(2)}_{L}&=& {1 \over 16~\pi^2} [ -2 Y_{L1}~(\gamma^{(1)}_{\phi_1}+
\gamma^{(1)}_{L^c})  
- 2 Y_{L2}~(\gamma^{(1)}_{\phi_2}+ \gamma^{(1)}_{L^c}) \nonumber \\
&&-({3
\over 2} \alpha_L + { 3 \over 4} \alpha_{B-L}) \gamma^{(1)}_{L}+
 {3 \over 2} b_L \alpha^2_L + { 3 \over 4}
b_{B-L} \alpha^2_{B-L}] 
, \\
\gamma^{(2)}_{L^c}&=& {1 \over 16~\pi^2} [ -2 Y_{L1}~(\gamma^{(1)}_{\phi_1}+
\gamma^{(1)}_{L})
- 2 Y_{L2}~(\gamma^{(1)}_{\phi_2}+ \gamma^{(1)}_{L})\nonumber \\
&&-({3                     
\over 2} \alpha_R + { 3 \over 4} \alpha_{B-L}) \gamma^{(1)}_{L^c}+  
 {3 \over 2} b_R \alpha^2_R + { 3 \over 4} 
b_{B-L} \alpha^2_{B-L}] 
, \\
\gamma^{(2)}_{\phi_i}&=& {1 \over 16~\pi^2} [ -Y_{Li}~(\gamma^{(1)}_{L}+ 
\gamma^{(1)}_{L^c})  - 3 Y_{Q_i}~(\gamma^{(1)}_{Q}+ \gamma^{(1)}_{Q^c})
\nonumber \\
&&- ({3
\over 2} \alpha_L 
+{3\over 2} \alpha_R) \gamma^{(1)}_{\phi_i}+
 {3 \over 2} b_L \alpha^2_L + { 3 \over 2}
b_{R} \alpha^2_{R}]. 
\eea
\subsection{Beta function coefficients for the gauge couplings}
The two loop RGE for the gauge couplings are given in general by Eqn
(\ref{2lrg}). The coefficients $b_i$, $b_{ij}$ and $a_{ik}$ depend on
the scale range considered, and are listed below:
\vskip .5cm
\noindent (A) For $M_Z \leq \mu \leq M_I$ the coefficients are:
\begin{equation}
b_i= \left(
\begin{array}{c} 2n_f+\frac{3}{5}n_d \\ -6+2n_f+n_d \\ -9+2 n_f
\end{array} \right)\, \;\;\; i=1,2,3 \,.
\end{equation}
\vskip .5cm \be b_{ij}= \left( \begin{array}{ccc}
\frac{38}{15}n_f+\frac{9}{25}\d & \frac{6}{5}\f+\frac{9}{5}\d &
\frac{88}{15}\f \\ \frac{2}{5}\f+\frac{3}{5}\d & -24+14\f+7\d & 8\f \\
\frac{11}{5}\f &3\f& -54+\frac{68}{3}\f \end{array} \right)\, \;\;\;
i,j= 1,2,3\,.
\end{equation}
\vskip .5cm \be a_{ik}= \left(\begin{array}{ccc} \frac{26}{5}&
\frac{14}{5} & \frac{18}{5} \\ 6 & 6 & 2 \\ 4 & 4 & 0
\end{array} \right) \,\;\;\; i=1,2,3\;, k=t,b,\tau \,.
\end{equation}
\vskip .5cm
\noindent(B) For $M_I \leq \mu \leq M_U$:
\vskip .5cm
\begin{equation}
b_i= \left(
\begin{array}{c} 2n_f+\frac{3}{2}(\l+\r) \\ -6+2n_f+\h+\l \\
-6+2\f+\h+\r \\ -9+2 n_f +3\c
\end{array} \right)\, \;\;\; i=B-L,\,L,\,R,\,3 \,.
\end{equation}
\vskip .5cm
{\small
\begin{eqnarray}
b_{ij}&= &\left(
\begin{array}{cccc}
\frac{7}{3} n_f+\frac{9}{4}(n_L+n_R) & 3 n_f+\frac{9}{2}n_L & 3
n_f+\frac{9}{2}n_R & \frac{8}{3} n_f \\ n_f+\frac{3}{2}n_L & -24+
14n_f+7(n_H+n_L) & 3n_H & 8n_f \\ n_f+\frac{3}{2}n_R & 3 n_H & -24+
14n_f+7(n_H+n_R) & 8n_f \\ \frac{1}{3}n_f & 3 n_f & 3 n_f & -54+
\frac{68}{3} n_f +34 n_c
\end{array} \right) \, \nonumber \\
& & i,j=B-L,\,L,\,R,\,3 \,. 
\end{eqnarray}
}
\vskip .5cm \be a_{ik} = \left(
\begin{array}{cccc}
2 & 2 & 6 & 6 \\ 12 & 12 & 4 & 4 \\ 12 & 12 & 4 & 4 \\ 8
& 8 & 0 & 0
\end{array} \right) \,\;\;\; i=B-L,\,L,\,R,\,3\,, \;\;k=Q1,\,Q2,\,L1,\,L2\,.
\end{equation}

\subsection{Proton decay parameters}
The proton decay formula is given by Eqn (\ref{pdecay}) \cite{hisano}:
\noindent where we have assumed the following values for the
parameters \cite{parameter}:
\begin{eqnarray*}
D&=&0.81~~;~~F=0.44~~;~~f_{\pi}= 139\, MeV \\ m_N&=&0.938\,GeV
~~;~~|V_{ud}|= 0.975~~;~~\alpha = 0.03 \, GeV^3 \,.
\end{eqnarray*}
\noindent The value of $\alpha$ is taken in such a way that proton
decay is maximal. In fact, its value has a range from 0.03 to 0.003 $GeV^3$.
\noindent The value of the renormalization factor $A_R$ varies from 2.7 
to 4.1 The values are plotted in as can be seen in Fig. (4). 
\begin{figure}[t] 
\epsfysize=9cm \epsfxsize=9cm \hfil \epsfbox{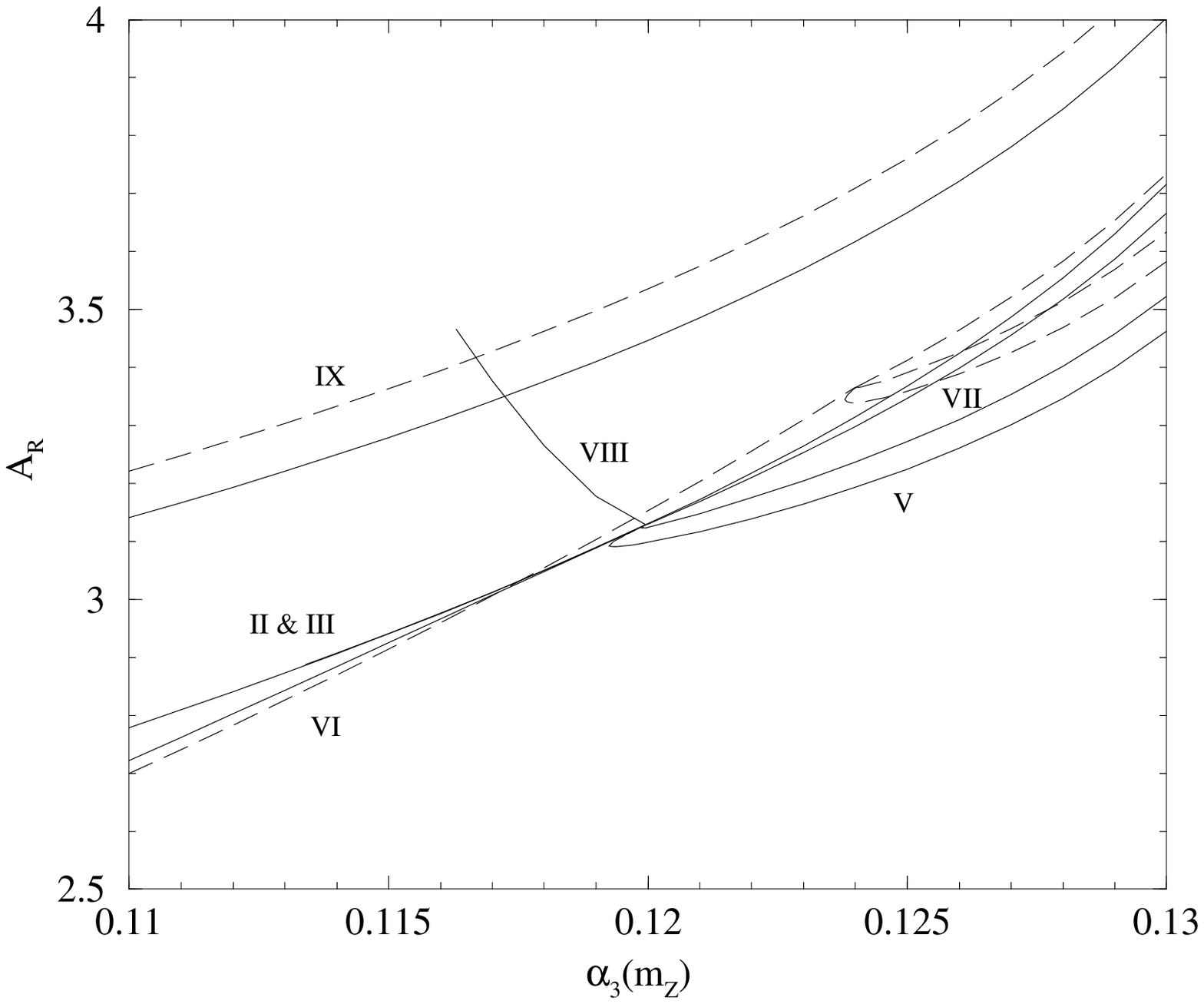} \hfil
\caption{ Variation of $A_R$ with respect to $\alpha_3(M_Z)$ in
different models. Solid and dashed lines are as before.}
\end{figure}

\subsection{The range of $\eta_b$}

Here we give the values of $\eta_b$ \cite{barger} used to extrapolate
$m_b(m_t)$ to $m_b(m_b)$, we have calculated its variation with
$\alpha_3(M_Z)$ for the different models studied in the text in Fig.
(5). We have taken $\a_2(M_Z)=0.3371$ and $\a_{1Y}(M_Z)=0.01696$. Our
calculation agrees with that of Ref. \cite{barger} quite well.

\begin{figure}[t]
\epsfysize=9cm \epsfxsize=9cm \hfil \epsfbox{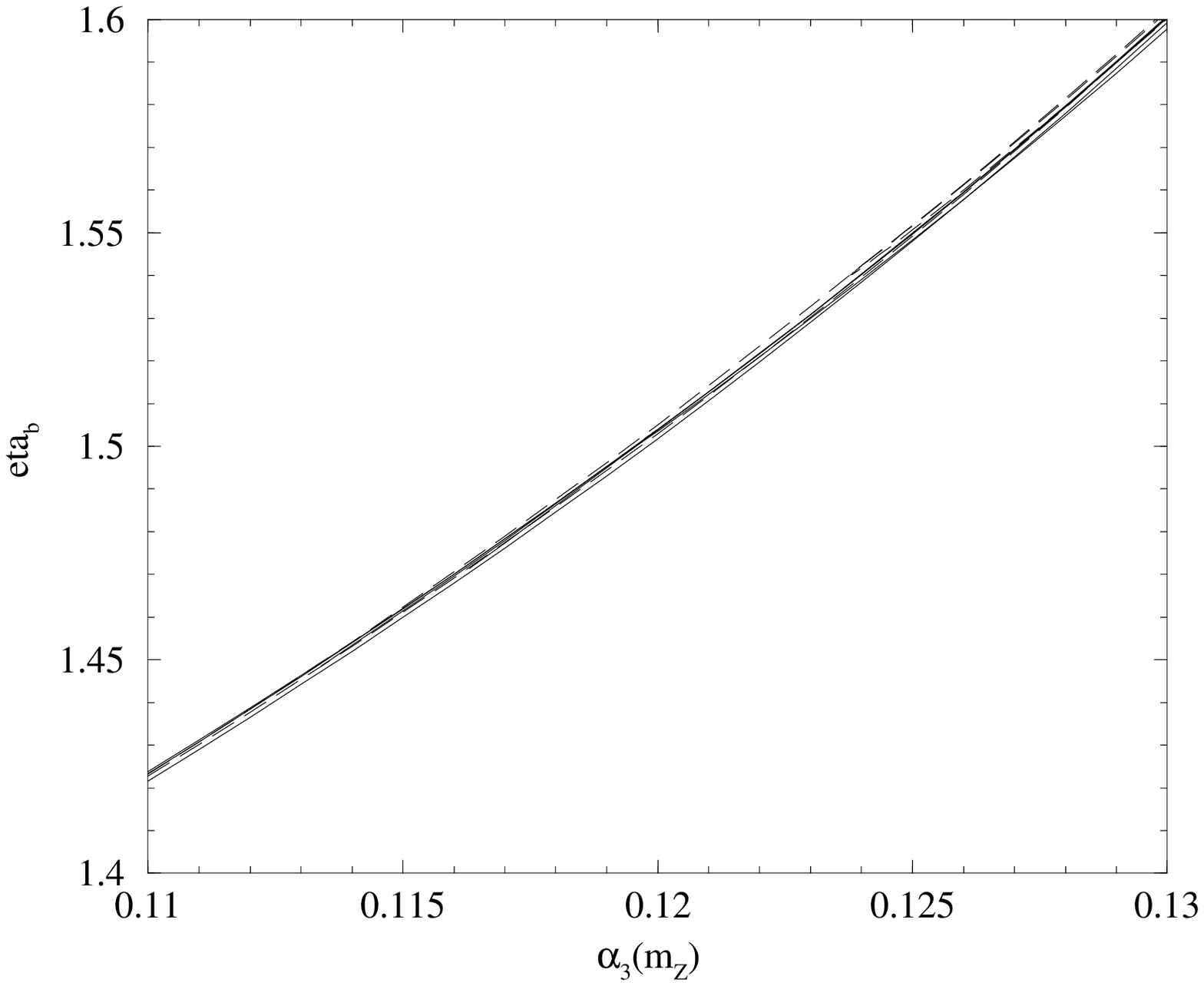} \hfil
\caption{ Variation of $\eta_b$ with respect to $\alpha_3(M_Z)$ in
different models. Solid and dashed lines are as before. }
\end{figure}

\end{document}